# 2-D Granular Model of Composite Elasticity using Molecular Dynamics Simulation

*Sparisoma Viridi, Widayani, and Siti Nurul Khotimah

Nuclear Physics and Biophysics Research Division
Institut Teknologi Bandung, Jalan Ganesha 10, Bandung 40132, Indonesia
*Email: dudung@fi.itb.ac.id

Abstract

Composite of two kinds of grains is modeled in two-dimension and the elasticity is calculated using molecular dynamics method implementing Gear predictor-corrector of fifth order. It has been observed that same composite concentration can be represented by several configurations of the two kinds of grains. Simulation results show a peak or maximum value of a parameter proportional to Young's modulus, which quantitatively agrees with reported experiment results. Ratio of $k_1/k_2$ used in simulation has influenced average value of the mentioned parameter monotonically, it is increasing as the ratio increasing.

**Introduction**

Elasticity of a composite depends nonlinearly on one of its compound concentration [1]. Granular model is proposed to explain the observation since this type of materials exhibits different elasticity behaviors during cycles of loading and unloading [2, 3] related to aging after and before the measurement cycles [4]. The correlation between Young's modulus and the confining pressure during tests is also reported [5]. In this work we propose a granular model for continuum media, as a contrary that is done by others [6]. A model for composite elasticity with a few granular gains will be conducted using molecular dynamics implementing Gear predictor-corrector algorithm [7].

**Simulation**

In this work it is assumed that the composite is build only from two kinds of compound, which are represented by two different types of grains. Then, two types of force are considered in this simulation, which are normal force and selective binding force. The fist type of force acts as repulsive force in order to prohibit two grains collapse into each other and the second acts as attractive force between two grains. The normal force has the formulation which is known as linear dash-pot model [8]

$$\vec{N}_{ij} = k_r \xi_{ij} \hat{r}_{ij} - k_v \dot{\xi}_{ij} \hat{r}_{ij}, \quad (1)$$

where $\xi_{ij}$ is defined as overlap between two grains and $\dot{\xi}_{ij}$ is the derivative of $\xi_{ij}$ with respect to time $t$. Overlap $\xi_{ij}$ is calculated through

$$\xi_{ij} = \max\left[0, \frac{1}{2}(d_i + d_j) - r_{ij}\right]. \quad (2)$$

And the selective binding force is defined as

$$\vec{B}_{ij} = \hat{r}_{ij}\left(k_B \frac{b_i b_j}{r_{ij}^2}\right), \quad (3)$$

with $b_i$ and $b_j$ have always integer positive value. Each type of grains has different value of $b$. Both Equations (1) and (3) used following definitions

$$\vec{r}_{ij} = \vec{r}_i - \vec{r}_j, \quad (4)$$

$$r_{ij} = |\vec{r}_{ij}| = \sqrt{\vec{r}_{ij} \cdot \vec{r}_{ij}}, \quad (5)$$

$$\hat{r}_{ij} = \frac{\vec{r}_{ij}}{r_{ij}}. \quad (6)$$

The binding force $\vec{B}_{ij}$ has a selective property, which means that grains with different types has different values of $k_B$ or explicitly it is chosen that

$$k_{ij} = \begin{cases} k_{AA}, & i = A, j = A \\ k_{AE}, & i = A, j = E \\ k_{AI}, & i = A, j = I \\ k_{IE}, & i = I, j = E \\ k_{II}, & i = I, j = I \\ 0, & \text{other } i, j \end{cases} = k_{ji}. \quad (7)$$

with the index A, E, and I stands for adhesive grains, ends grains, and interstitial grains, respectively. Illustration about the model is given in Figure 1.

Gear predictor-corrector algorithm of fifth order [7] is chosen in the molecular dynamics method used in the simulation, which has two steps: prediction step (written with upper index $p$) and correction step for every particular grain. The first step is formulated as

$$\begin{pmatrix} \vec{r}_0^{\,p}(t+\Delta t) \\ \vec{r}_1^{\,p}(t+\Delta t) \\ \vec{r}_2^{\,p}(t+\Delta t) \\ \vec{r}_3^{\,p}(t+\Delta t) \\ \vec{r}_4^{\,p}(t+\Delta t) \\ \vec{r}_5^{\,p}(t+\Delta t) \end{pmatrix} = \begin{pmatrix} 1 & 1 & 1 & 1 & 1 & 1 \\ 0 & 1 & 2 & 3 & 4 & 5 \\ 0 & 0 & 1 & 3 & 6 & 10 \\ 0 & 0 & 0 & 1 & 4 & 10 \\ 0 & 0 & 0 & 0 & 1 & 5 \\ 0 & 0 & 0 & 0 & 0 & 1 \end{pmatrix} \begin{pmatrix} \vec{r}_0(t) \\ \vec{r}_1(t) \\ \vec{r}_2(t) \\ \vec{r}_3(t) \\ \vec{r}_4(t) \\ \vec{r}_5(t) \end{pmatrix}. \quad (8)$$

And the correction step will give the corrected value of $\vec{r}_n(t+\Delta t)$ through

$$\begin{pmatrix} \vec{r}_0(t+\Delta t) \\ \vec{r}_1(t+\Delta t) \\ \vec{r}_2(t+\Delta t) \\ \vec{r}_3(t+\Delta t) \\ \vec{r}_4(t+\Delta t) \\ \vec{r}_5(t+\Delta t) \end{pmatrix} = \begin{pmatrix} \vec{r}_0^{\,p}(t+\Delta t) \\ \vec{r}_1^{\,p}(t+\Delta t) \\ \vec{r}_2^{\,p}(t+\Delta t) \\ \vec{r}_3^{\,p}(t+\Delta t) \\ \vec{r}_4^{\,p}(t+\Delta t) \\ \vec{r}_5^{\,p}(t+\Delta t) \end{pmatrix} + \begin{pmatrix} c_0 \\ c_1 \\ c_2 \\ c_3 \\ c_4 \\ c_5 \end{pmatrix} \Delta \vec{r}_2(t+\Delta t), \quad (9)$$

with

$$\Delta \vec{r}_2(t+\Delta t) = \vec{r}_2(t+\Delta t) - \vec{r}_2^{\,p}(t+\Delta t). \quad (10)$$

The term $\vec{r}_n(t+\Delta t)$ is defined as

$$\vec{r}_n(t) = \frac{(\Delta t)^n}{n!}\left[\frac{d^n \vec{r}_0(t)}{dt^n}\right], \quad (11)$$

where $\vec{r}_0$ is position of a grain. The term $\vec{r}_2(t+\Delta t)$ in correction term in Equation (10) is obtained from Newton's second law of motion. For example, particle $i$ has

$$[\vec{r}_2(t+\Delta t)]_i = \frac{(\Delta t)^2}{m_i} \times \sum_{j \neq i} \vec{B}_{ij}(t+\Delta t) + \vec{N}_{ij}(t+\Delta t). \quad (12)$$

The left part of Equation (12) is calculated using $\vec{r}_n^{\,p}(t+\Delta t)$.

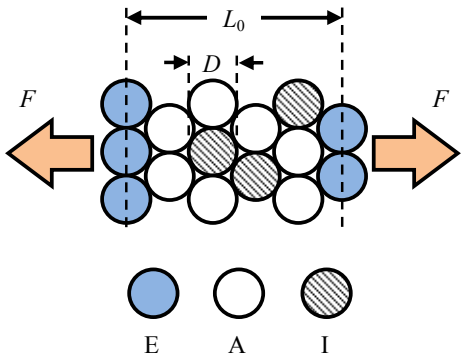

Figure 1. Granular model of a 2-D bulk materials with equal grains diameter $D$, initial length $L_0$, applied force $F$, for three kinds of grains: ends grains (E), adhesive grains (A), and interstitial grains (I).

Each kind of grains play different role. Ends gains (E) are always aligned as they are moved and used to calculate length of the model. Adhesive grains (A) can bind ends grains, other adhesive grains, and interstitial grains (I). There is no attractive interaction between interstitial grains. For simplicity the end grains will be chosen the same as the adhesive grains.

Concentration of interstitial grains $C$ is calculated using

$$C = \frac{N_I}{N_A + N_I} \quad (13)$$

with $N_A$ and $N_I$ are number of adhesive grains (A) and interstitial grains (I), respectively. Variation of placements of interstitial grains for the same concentration $C$ has not yet been conducted as proposed in [9, 10] using Gaussian sequences, but only selected configuration is considered in this work.

Instead of varying the applied force $F$, the length $L$ is set to some certain value and after the system given in Figure 1 achieves its equilibrium the force suffered by ends grains $F$ is calculated. From sets of value of $(L, F)$ the Young modulus of the bulk materials $Y$ can be determined through

$$L = L_0 + \left(\frac{L_0}{A_0 Y}\right) F. \quad (14)$$

In this case $A_0$ is about $2.5 D^2$. It can be found directly from common linear regression formula that

$$Y = \left(\frac{L_0}{A_0}\right) \frac{\sum_{k=1}^{M}(F_k - \langle F \rangle)^2}{\sum_{k=1}^{M}(F_k - \langle F \rangle)(L_k - \langle L \rangle)}, \quad (15)$$

with

$$\langle X \rangle = \frac{1}{N} \sum_{k=1}^{M} X_k. \quad (16)$$

The variable $X$ can be $L$ or $F$ with number of data is $M$. The data, set of $(L, F)$, is produced using the simulation.

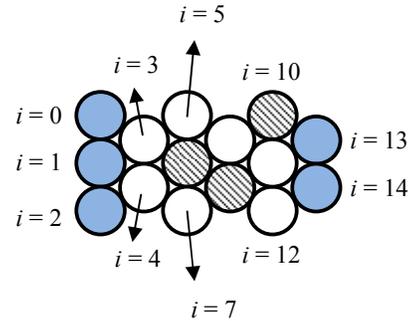

Figure 2. Initial configuration of all grains as determined using Equation (17) and (18).

### Results and discussion

For the initial configuration following formulas are used to determine all grains initial position

$$p_i = [(i \operatorname{div} 5) + i] \operatorname{div} 3, \quad (17)$$

$$x_i = x_0 + p_i D \cos\left(\frac{\pi}{6}\right), \quad (18)$$

$$q_i = [1 - (p_i \bmod 2)][1 - (i \bmod 5)] + (p_i \bmod 2)[3.5 - (i \bmod 5)], \quad (19)$$

$$y_i = y_0 + 2q_i D \sin\left(\frac{\pi}{6}\right), \quad (20)$$

which are illustrated in Figure 2 for number of grains $N = 15$. Position of grains $i = 1$ is defined as $(x_0, y_0)$. In Equation (17) and (19) there are two integer operators, which are div and mod. The first is to find the integer part from division result and the second is to find the remaining (modulo) of integer division. As result of Equation (17)-(20) is given in Figure 3 for $D = 1$ and $N = 15$, which gives $L_0$ about 4.33.

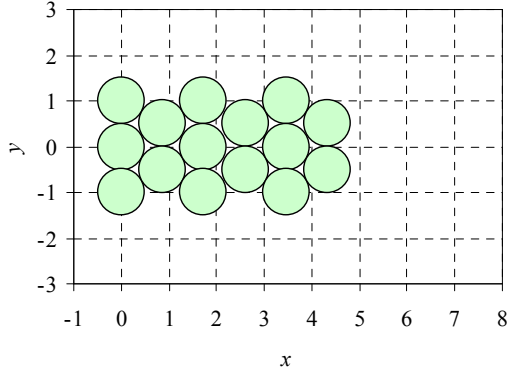

Figure 3. Implementation of Equation (17)-(20) for $D = 1$ and $N = 15$ in producing initial configuration.

Following value of parameters are used in the simulation $N = 15$, $D = 1$, $x_0 = 0$, $y_0 = 0$, $m = 1$, $k_{AA} = k_{AI} = k_{AE} = k_{EI} = k_1 = 10^{-2}$, $k_{II} = k_2 = 10^{-5}$, $b = 1$, $k_r = 50$, $k_v = 0.5$, $t_i = 0$, $t_f = 50$, $T_s = 0$, $\Delta t = 10^{-3}$.

Several concentrations of 0, 0.2, 0.4, and 0.6 are calculated. For $C = 0.2$ several configurations are calculated. Plot of parameter proportional to Young's modulus $Y_1$ against concentration of interstitial grains $C$ is given in Figure 4.

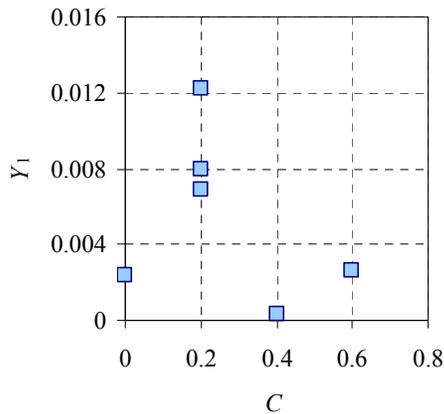

Figure 4. Parameter proportional to Young's modulus $Y_1$ is influenced by interstitial grains concentration $C$.

It can be seen from Figure 4 that even for same value of $C$ it can produce different value of $Y_1$. Position of adhesive grains and interstitial grains for results shown in Figure 4 is given in Table 1 with 0 indicates adhesive grains and 1 indicates interstitial grains. Position of grains obeys the rules given previously in Equation (17)-(20) or in Figure 2.

Table 1. Concentration $C$, grains configuration, and its parameter proportional to Young's modulus $Y_1$.

| $C$ | $g_0 g_1 \cdots g_{13} g_{14}$ | $Y_1$ |
|---|---|---|
| 0   | 000 00 000 00 000 00 | 0.0024 |
| 0.2 | 000 10 100 10 000 00 | 0.008  |
| 0.2 | 000 10 011 10 000 00 | 0.0122 |
| 0.2 | 000 10 001 10 000 00 | 0.0069 |
| 0.4 | 000 10 101 11 100 00 | 0.0003 |
| 0.6 | 000 10 111 11 111 00 | 0.0026 |

Result in Figure 4 shows a nearly similar result when they are compared to the reported experiment results [1], that there is a peak or maximum value of Young's modulus as the concentration of interstitial grains increasing. Unfortunately, both results can not be compared directly since there is a lack of scaling information about how to map simulation result to the experiment one.

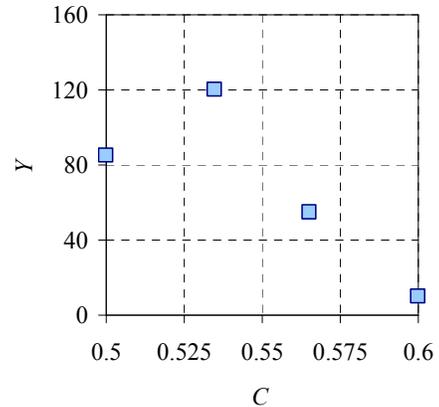

Figure 5. Experiment results reported in [1] as comparison to simulation results.

From Figure 1, when all the parameters used in simulation are already interpreted correctly to its physical meaning, the simulation results can be used to study what is the real influence of concentration $C$ to value of Young's modulus related to the grains configuration in microscale. Then the degree of mixture can also be calculated. The "degeneracy" of grains configuration that different configurations can produce same Young's modulus is an interesting phenomenon. Further study can be addressed to this aspect.

The influence of ratio $k_1/k_2$ to the parameter $Y_1$ is also investigated and the trends are a little bit unpredictable as shown in Figure 6 but its average, $\langle Y_1 \rangle$, is increasing as value of $k_1/k_2$ increasing. This relation is given in Figure 7.

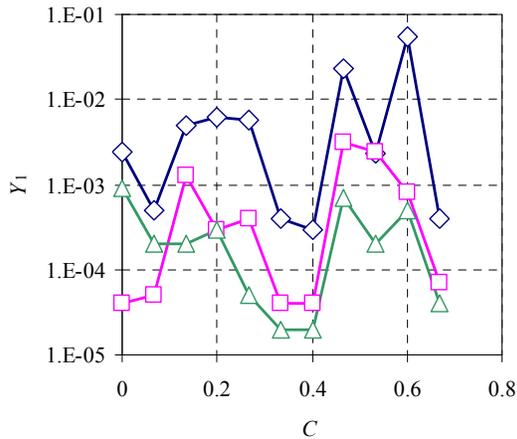

Figure 6. Simulation result for parameters proportional to Young's modulus $Y_1$ as function of concentration $C$ for different values of $k_1/k_2$: 10 (△), 100 (□), 1000 (◊).

Various concentrations $C$ in Figure 6 uses following configuration of $g_0 g_1 \cdots g_{13} g_{14}$, which are 000 00 000 00 000 00, 000 10 000 00 000 00, 000 11 000 00 000 00, .., 000 11 111 11 111 00. This configuration gives value of $C$: 0.000, 0.067, 0.133, 0.200, 0.267, 0.333, 0.400, 0.467, 0.533, 0.600, and 0.667, respectively.

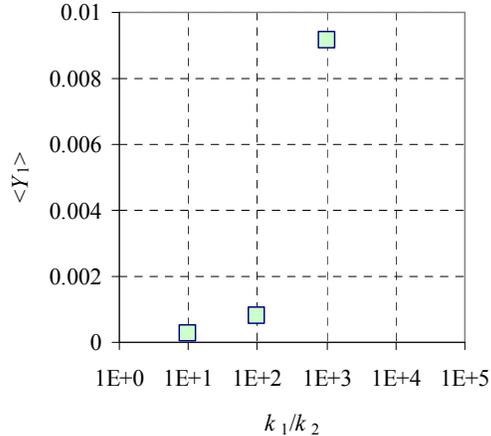

Figure 7. Influence of $k_1/k_2$ ratio to average value of parameter proportional to Young's modulus $Y_1$.

Ratio of binding force of adhesive grains to force of interstitial grains, which is represented by $k_1/k_2$, plays important role in determining average value of $Y_1$ as seen in Figure 7. This information can be used to "tune" Young's modulus of a composite by varying its composition. But this idea must be supported by mechanical information, e.g. yield strength or ultimate strength, which is out of scope of this work, since obtaining correct $Y_1$ for a composite but reducing its yield or ultimate strength is not what we want to get, isn't it?

## Conclusion

It can be concluded that the proposed model can produced a maximum value of Young's modulus as reported in other work. Direct comparison needs scaling information that has not yet been defined. Average value of parameter proportional to Young's modulus $\langle Y_1 \rangle$ is increasing as $k_1/k_2$ increasing.

## Acknowledgements

Authors would like to thanks ITB for supporting this work through Hibah Peningkatan Kapasitas in 2010-2011.